\documentclass[conference]{IEEEtran}
\IEEEoverridecommandlockouts

\usepackage{cite}
\usepackage{amsmath,amssymb,amsfonts}
\usepackage{graphicx}
\usepackage{textcomp}
\usepackage{xcolor}
\usepackage{array}
\usepackage{booktabs}
\usepackage{makecell}
\usepackage{cuted}
\usepackage{capt-of}

\newcolumntype{P}[1]{>{\centering\arraybackslash}p{#1}}

\def\BibTeX{{\rm B\kern-.05em{\sc i\kern-.025em b}\kern-.08em
    T\kern-.1667em\lower.7ex\hbox{E}\kern-.125emX}}

\begin{document}

\title{Hardware-in-the-Loop Phase-Aware CNN for Real-Time 5G Channel Estimation}

\author{
\IEEEauthorblockN{
J.~Zolfaghari-Bengar,
R.~Rony,
E.~Gomez-de-Lope,
A.~Villena-Rodriguez,
A.~Mahadevan,
N.~Kourtellis
}
\IEEEauthorblockA{
Keysight AI Labs\\
\{javad.z, rakibul.rony, elisa.gomez-de-lope, alejandro.villena-rodriguez, abhinav.mahadevan, nicolas.kourtellis\}@keysight.com}}

\maketitle

\begin{strip}
\centering
\includegraphics[width=\textwidth]{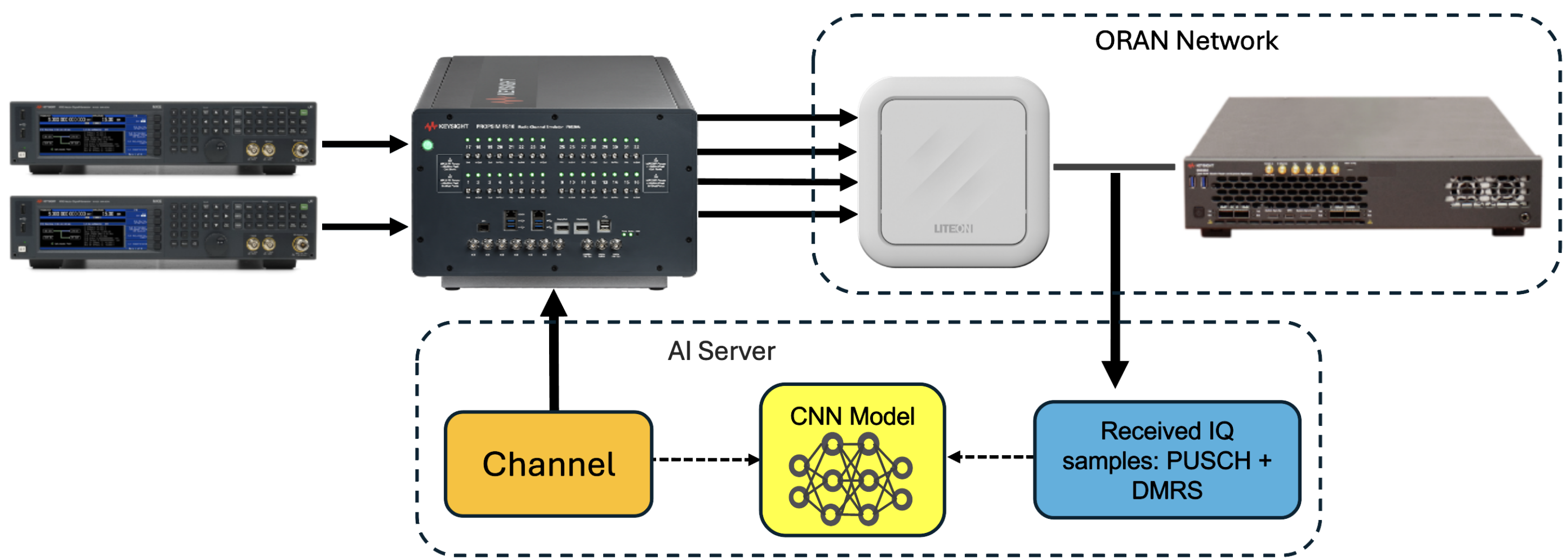}
\captionof{figure}{Hardware-in-the-loop platform used for collecting the 5G uplink channel-estimation dataset, integrating two RF generators, a PROPSIM channel emulator, an O-RAN RU, DU emulation, and an AI server. The conference demo performs real-time CNN inference and visualization using the captured hardware-derived data.}
\label{fig:testbed}
\end{strip}

\begin{abstract}
This demo presents real-time AI-based uplink channel-estimation inference using data collected from a hardware-in-the-loop 5G platform. The data-collection setup integrates commercial RF signal generation, programmable channel emulation, an O-RAN Radio Unit, DU emulation, and a lightweight phase-aware convolutional neural network (CNN) that estimates the channel response directly from received DMRS signals. Unlike simulation-only evaluations, the hardware-derived dataset exposes the estimator to practical RF and system-level impairments, including calibration mismatches, synchronization imperfections, quantization effects, phase noise, and implementation-specific nonlinearities. During the demo, attendees will observe real-time CNN inference and channel reconstruction using captured hardware-generated DMRS observations and compare the proposed CNN against Least Squares (LS) and frequency-domain LMMSE baselines. The objective is to showcase a practical AI-native physical-layer inference pipeline that combines hardware-derived 5G data with real-time neural channel estimation for future 5G-Advanced and 6G systems.
\end{abstract}


\section{Introduction}

Accurate channel estimation is essential for reliable demodulation, interference management, and beamforming in 5G and emerging 6G systems. As deployments become denser and channels more dynamic, received pilot signals are increasingly affected not only by propagation and noise, but also by RF impairments, synchronization behavior, radio-unit processing, and baseband implementation details~\cite{b1}.

Classical estimators remain useful baselines but have practical limitations. Least Squares (LS) is simple and fast, yet it directly propagates noise and does not exploit channel statistics. MMSE/LMMSE estimators can improve accuracy when reliable covariance information is available, but such statistics are often unavailable or difficult to maintain in dynamic hardware-impaired environments~\cite{b2}. Deep learning (DL)-based estimators have shown strong potential for OFDM and MIMO channel estimation~\cite{b3,b4,b5,b6}, but still face challenges such as phase wrapping, limited generalization across UE and antenna configurations~\cite{b7,b8}, and deployment constraints.

The complete data-collection and inference workflow is shown in Fig.~\ref{fig:testbed}. This demo presents an AI inference pipeline for uplink channel estimation using received DMRS signals collected from a hardware-in-the-loop platform. The demo provides a practical example of AI-native physical-layer processing for future O-RAN and 6G systems~\cite{b9}. The data-collection system combines commercial RF signal generation, programmable channel emulation, O-RAN radio processing, and DU emulation. A lightweight phase-aware CNN reconstructs the channel magnitude and phase using sine--cosine phase encoding. During the conference demo, the captured hardware-derived data are processed by the CNN in real time and compared with LS and frequency-domain LMMSE estimators.

\section{Demonstration Platform}

As shown in Fig.~\ref{fig:testbed}, the testbed used for data collection emulates two independent single-antenna UEs using Keysight MXG N5182B RF generators. The generated uplink DMRS signals passed through a Keysight PROPSIM F8800B channel emulator configured with controlled 5G propagation scenarios and were then fed into a Liteon O-RAN Radio Unit for RF-to-baseband conversion. A Keysight S5040A DU emulator processed the resulting baseband I/Q samples through the 5G uplink chain. The received PUSCH and DMRS I/Q samples, together with the corresponding channel responses, were captured and forwarded to an AI compute server for model training, evaluation, inference, and visualization. The conference demo uses these captured hardware-derived samples to perform real-time CNN inference without requiring the complete RF testbed at the venue. The data-collection setup used a $2\times4$ MIMO uplink configuration with two UEs and a four-antenna receiver. The main 5G NR parameters are summarized in Table~\ref{tab:sys_params}, including a $3.5$~GHz carrier, $100$~MHz bandwidth, $30$~kHz subcarrier spacing, $4096$-point FFT, $273$ resource blocks, $1638$ evaluated subcarriers, and DMRS Configuration Type~I.

\begin{table}[t]
\caption{Hardware Data-Collection and AI Inference Platform}
\label{tab:demo_platform}
\centering
\small
\renewcommand{\arraystretch}{1.12}
\begin{tabular}{p{2.65cm}p{4.55cm}}
\toprule
\textbf{Component} & \textbf{Description} \\
\midrule
RF generators & $2\times$ Keysight MXG N5182B \\
Channel emulator & Keysight PROPSIM F8800B \\
Radio Unit & Liteon O-RAN RU \\
DU emulator & Keysight S5040A \\
Training platform & Apple MacBook Pro (M4 Pro, 24 GB unified memory) \\
GPU acceleration & Apple integrated GPU using TensorFlow Metal \\
Inference latency & $\approx 30$ ms per batch ($B=32$, unoptimized implementation) \\
\bottomrule
\end{tabular}
\end{table}

\begin{table}[t]
\caption{5G NR Uplink System Parameters}
\label{tab:sys_params}
\centering
\small
\renewcommand{\arraystretch}{1.15}
\begin{tabular}{p{3.4cm}p{4cm}}
\toprule
\textbf{Parameter} & \textbf{Value} \\
\midrule
Wireless standard & 5G NR uplink \\
Carrier frequency & $3.5$~GHz \\
Bandwidth & $100$~MHz \\
Numerology / SCS & $1$ / $30$~kHz \\
FFT size & $4096$ \\
Resource blocks & $273$ \\
Evaluated subcarriers & $1638$ \\
DMRS config. / ports & Type~I / $[0,2]$ \\
Delay spread & $300$~ns \\
MIMO configuration & $2\times4$ uplink (2 UEs, 4 Rx) \\
Channel scenarios & RMa, UMa, UMi (LOS/NLOS) \\
\bottomrule
\end{tabular}
\end{table}

Table~\ref{tab:demo_platform} summarizes the main hardware components used for data collection and the AI platform used for training and real-time inference, while Table~\ref{tab:sys_params} lists the 5G~NR uplink parameters used during signal generation and channel emulation. The hardware data-collection platform exposes the learning model to realistic RF effects that are difficult to capture with pure software simulation, including calibration offsets, quantization effects, synchronization imperfections, phase noise, and hardware nonlinearities.

\section{AI Processing Pipeline}

\begin{figure}[t]
\centering
\includegraphics[width=\columnwidth]{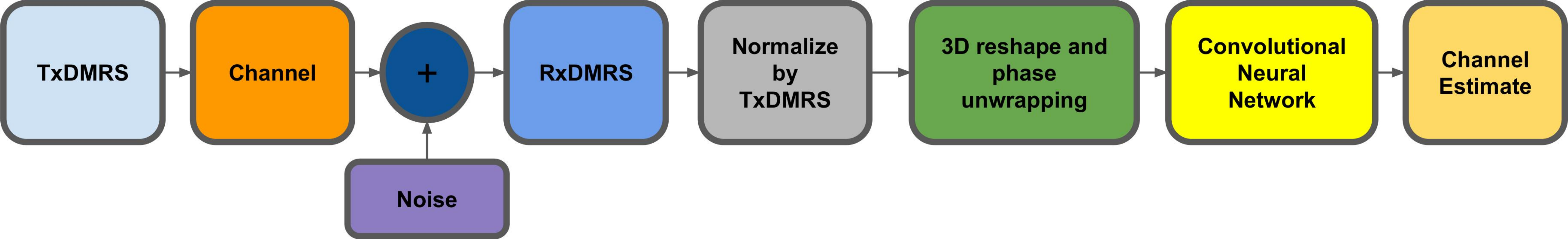}
\caption{AI processing pipeline used in the demonstration: previously captured hardware-derived DMRS observations are normalized by the transmitted DMRS, reshaped with sine--cosine phase encoding, and processed by the CNN in real time to produce the channel estimate.}
\label{fig:pipeline}
\end{figure}

The CNN operates on captured received DMRS observations. Each complex received DMRS value is represented using its magnitude and phase encoded as sine and cosine components:
\begin{equation}
\mathbf{x}_{\mathrm{in}} = \left[|Y_{\mathrm{DMRS}}|,\; \sin(\angle Y_{\mathrm{DMRS}}),\; \cos(\angle Y_{\mathrm{DMRS}})\right].
\end{equation}
The network predicts the channel response using a corresponding channel representation,
\begin{equation}
\mathbf{x}_{\mathrm{out}} = \left[|H|,\; \sin(\angle H),\; \cos(\angle H)\right].
\end{equation}
During inference, the estimated phase is recovered using
\begin{equation}
\widehat{\angle H} = \operatorname{atan2}\left(\widehat{\sin(\angle H)},\widehat{\cos(\angle H)}\right).
\end{equation}
This representation avoids the artificial discontinuity of wrapped phase at $\pm\pi$ and enables stable regression of both magnitude and phase. The overall preprocessing and real-time inference flow, from captured DMRS observations through normalization and CNN-based reconstruction, is illustrated in Fig.~\ref{fig:pipeline}.

\begin{table}[t]
\caption{CNN Architecture Used for Channel Estimation}
\label{tab:cnn_arch}
\centering
\small
\renewcommand{\arraystretch}{1.15}
\begin{tabular}{lccc}
\toprule
\textbf{Layer} & \textbf{Type} & \textbf{Filters} & \textbf{Kernel / Act.} \\
\midrule
Input & --- & --- & $[\,\cdot,\, N_{sc},\, U\!\times\!R,\, 3\,]$ \\
Layer 1 & Conv2D + BN & $32$ & $(5,1)$ / ReLU \\
Layer 2 & Conv2D + BN & $64$ & $(5,3)$ / ReLU \\
Layer 3 & Conv2D + BN & $64$ & $(3,3)$ / ReLU \\
Layer 4 & Conv2D + BN & $32$ & $(3,1)$ / ReLU \\
Layer 5 & Conv2D & $3$ & $(1,1)$ / Linear \\
\bottomrule
\end{tabular}
\end{table}

The CNN is lightweight and composed of five 2D convolutional layers with batch normalization, detailed in Table~\ref{tab:cnn_arch}, that learn spatial and frequency-domain correlations across receive antennas, UEs, and subcarriers. The model has a footprint under $1$~MB, enabling real-time deployment on edge hardware, and is trained with the Adam optimizer (learning rate $0.001$, batch size $32$) for $50$ epochs. It is compared against two baselines: 1)~LS, computed by dividing the received DMRS by the transmitted DMRS per subcarrier, receive antenna, and UE; 2)~an oracle frequency-domain LMMSE reference formed from LS channel vectors using an empirical frequency covariance matrix.

\section{Demo Scenario and Evaluation Setup}

The hardware-in-the-loop platform was used to generate and validate the dataset across nine representative 3GPP scenarios and to evaluate channel reconstruction accuracy, inference latency, and throughput.

During the conference demo, the captured hardware-derived DMRS observations are provided to the inference application, which performs CNN-based channel reconstruction and visualization in real time. Attendees will be able to observe the estimated channel magnitude and phase and compare the CNN output with the LS and frequency-domain LMMSE baselines.

A key aspect of the evaluation is that the dataset was generated by the hardware testbed of Fig.~\ref{fig:testbed}, rather than by pure software simulation. The hardware-derived traces span the nine representative 3GPP scenarios listed in Table~\ref{tab:dataset}, covering rural macro (RMa), urban macro (UMa), and urban micro (UMi) environments under indoor, outdoor LOS, and outdoor NLOS conditions. For each scenario, $12{,}000$ samples were used for training, $1{,}000$ for validation, and $2{,}000$ for testing, resulting in $135{,}000$ hardware-derived samples in total. Each sample comprises the transmitted DMRS, the received DMRS, and the corresponding ground-truth channel frequency response.

\begin{table}[t]
\caption{Channel Scenarios in the Demonstration Dataset}
\label{tab:dataset}
\centering
\small
\renewcommand{\arraystretch}{1.15}
\begin{tabular}{lccc}
\toprule
\textbf{Scenario} & \textbf{Model} & \textbf{Indoor/Outdoor} & \textbf{LOS/NLOS} \\
\midrule
RMa\_indoor & RMa & Indoor & --- \\
RMa\_LOS & RMa & Outdoor & LOS \\
RMa\_NLOS & RMa & Outdoor & NLOS \\
UMa\_indoor & UMa & Indoor & --- \\
UMa\_LOS & UMa & Outdoor & LOS \\
UMa\_NLOS & UMa & Outdoor & NLOS \\
UMi\_indoor & UMi & Indoor & --- \\
UMi\_LOS & UMi & Outdoor & LOS \\
UMi\_NLOS & UMi & Outdoor & NLOS \\
\bottomrule
\end{tabular}
\end{table}

\section{Representative Results}

\begin{figure}[t]
\centering
\includegraphics[width=\columnwidth]{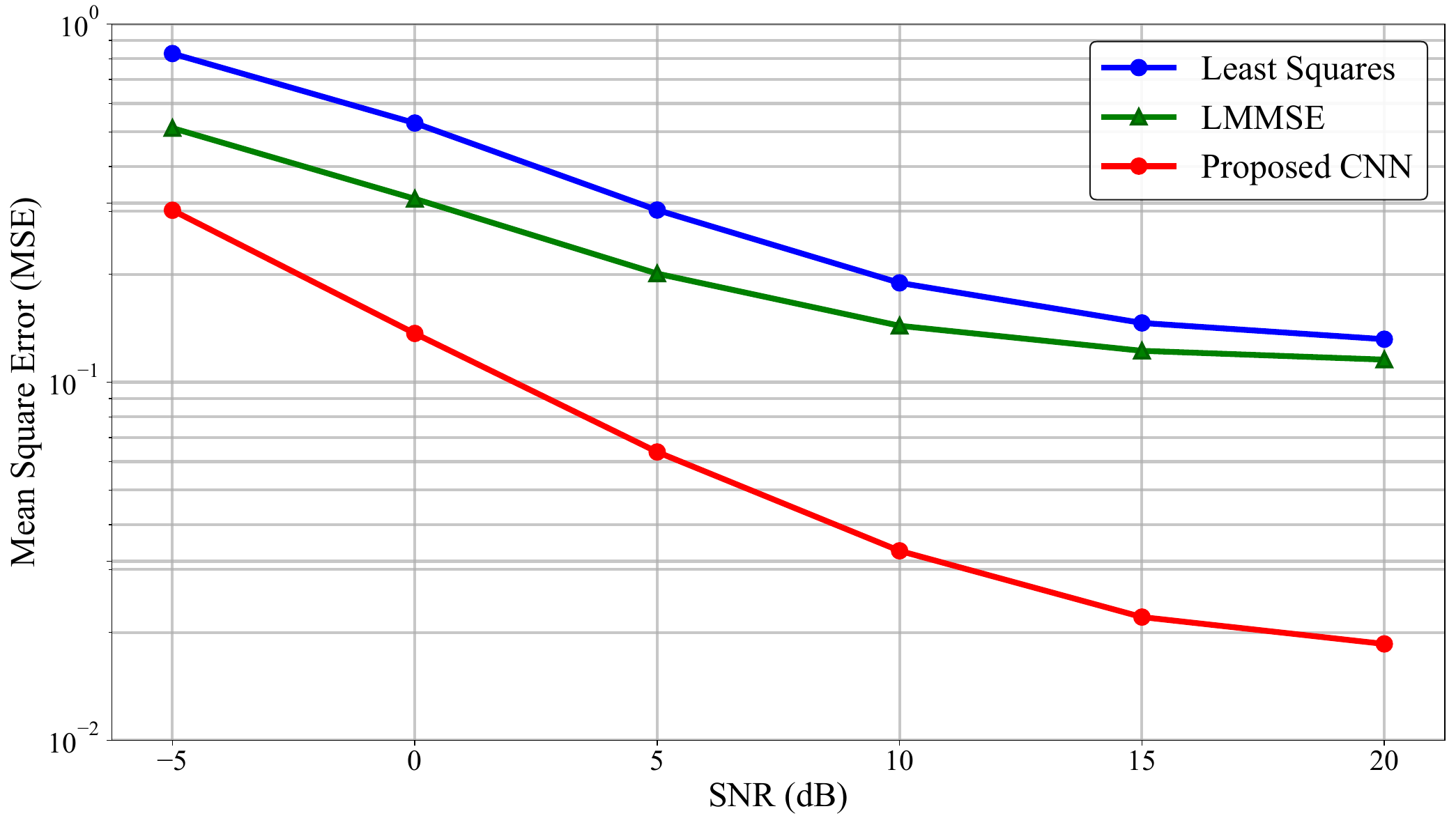}
\caption{Channel-estimation error versus SNR for LS, LMMSE, and the proposed phase-aware CNN on hardware-derived data.}
\label{fig:snr}
\end{figure}

Fig.~\ref{fig:snr} reports a representative evaluation of the proposed CNN, LS, and oracle frequency-domain LMMSE estimators across an SNR sweep using hardware-derived channel realizations. The phase-aware CNN achieves the lowest MSE across the evaluated SNR range, with the largest gains observed at low and moderate SNR values where noise has the strongest impact.

Aggregated over the test set, the CNN reduces the magnitude MSE against the LS baseline from $0.0860$ to $0.0105$, corresponding to nearly an order-of-magnitude improvement. In the phase domain, the CNN reduces the MSE from $1.9554$ to $0.7822$, indicating smoother and more accurate reconstruction of both magnitude and phase.

\section{Conclusion}

This demo presents real-time AI-based channel-estimation inference using data collected from a hardware-in-the-loop 5G uplink platform. The hardware setup combines commercial RF equipment, O-RAN components, and controlled channel emulation to produce realistic channel-estimation data. During the conference demo, attendees can inspect the CNN inference and visualization process, compare the CNN with classical baselines, and observe real-time channel reconstruction using the captured hardware-derived observations.


\begin{thebibliography}{00}

\bibitem{b1}
B. Li, Q. Zheng, X. Tian, M. Yang, G. Gui, W. Jiang, H. Lei, J. Jiang, F. Shu, A. Elhanashi, and S. Saponara,
``A Survey of Artificial Intelligence Enabled Channel Estimation Methods: Recent Advance, Performance, and Outlook,'' 2025.

\bibitem{b2}
G. Bacci, A. A. D'Amico, and L. Sanguinetti,
``Low-complexity MMSE channel estimation for wideband massive MIMO systems,'' 2024.

\bibitem{b3}
H. Ye, G. Y. Li, and B.-H. Juang,
``Power of Deep Learning for Channel Estimation and Signal Detection in OFDM Systems,'' 2018.

\bibitem{b4}
H. He, C.-K. Wen, S. Jin, and G. Y. Li,
``Deep Learning-Based Channel Estimation for Beamspace mmWave Massive MIMO Systems,'' 2018.

\bibitem{b5}
M. Soltani, V. Pourahmadi, A. Mirzaei, and H. Sheikhzadeh,
``Deep Learning-Based Channel Estimation,'' 2019.

\bibitem{b6}
E. G{\'o}mez-de-Lope, J. Zolfaghari-Bengar, R. Rony, A. Mahadevan, and N. Kourtellis,
``X-CE: Scenario-Aware Explainable Deep Learning for MIMO Channel Estimation,''
in \textit{Proc. IEEE INFOCOM 2026 -- IEEE Conference on Computer Communications}, 2026, pp. 1--7,
doi: 10.1109/INFOCOM59046.2026.11571707.


\bibitem{b7}
Z. Jiang, S. Chen, A. F. Molisch, and R. Vannithamby,
``Exploiting Wireless Channel State Information Structures Beyond Linear Correlations: A Deep Learning Approach,'' 2019.

\bibitem{b8}
S. Sibio, C. Sestito, S. B. Smida, Y. Ding, and G. Goussetis,
``Low-Complexity Convolutional Neural Network for Channel Estimation,'' 2024.

\bibitem{b9}
A. K. Gizzini and M. Chafii,
``Deep Learning Based Channel Estimation in High Mobility Communications Using Bi-RNN Networks,'' 2023.









\end{thebibliography}
\end{document}